\documentclass[pra,bibnotes,twocolumn]{revtex4}
\usepackage{graphicx}
\begin{document}
\draft
\def\ds{\displaystyle}
\title{  Robust bulk states }
\author{C. Yuce}
\address{Department of Physics, Eskisehir Technical University, Turkey }
\email{cyuce@anadolu.edu.tr}
\date{\today}
\begin{abstract}
We predict topologically robust zero energy bulk states in a disordered tight binding lattice. We explore a new kind of order and discuss that zero energy states exist in a system iff its Hamiltonian is noninvertible. We show that they are robust against any kind of disorder as long as the disordered Hamiltonian is noninvertible, too.
\end{abstract}
\maketitle
\section{Introduction}
A topological insulator (TI) is an insulator in its interior but has conducting states at its surface. This means that particles can only move along the surface of the topological insulating material. A topological insulator has gapless edge states that are immune to symmetry protecting disorder. The topological concept dates back to 1980 when the quantum Hall effect in two-dimensional (2D) electron gas at the interface of a semiconductor heterojunction under a strong external magnetic field was observed \cite{klitzing}. Thouless, Kohmoto, Nightingale and Nijs (TKNN) had a brilliant idea \cite{thoules}. They said that this effect is in fact a topological effect. Haldane proposed a model to realize topologically nontrivial system without strong magnetic field, which is necessary in the quantum Hall effect to break time-reversal symmetry \cite{halda}. Several years later, Kane and Mele made a breakthrough in the study of topological insulators \cite{kane1,kane2}. They found a new topological number: $Z_2$ number. CdTe/HgTe/CdTe quantum well was predicted to be a first 2D topological insulator \cite{bernevig} and this structure was experimentally realized soon after \cite{konig}. In years, many other topological systems such as topological crystalline insulators protected by crystal symmetries have been introduced \cite{andofu}. Recently, $n$th-order topological insulators that have $(d-n)$D gapless boundary states in $d$ dimension was explored \cite{hoti1,hoti2}. We emphasize that topological phase does not only exist in Hermitian systems but also in non-Hermitian systems. In recent years, non-Hermitian extension of topological insulators and superconductors have attracted considerable attention \cite{nonh1,nonh2,nonh3,nonh4,nonh5,nonh6}. The theory of topological insulators and superconductors has not been fully constructed. Fortunately, great progress has been made in our understanding of these systems. We refer the reader for details to some excellent review papers on this subject \cite{rev01,rev02,rev03,rev04}. \\
One of the most significant features of topological insulators is the bulk-boundary correspondence. According to that correspondence, there exist gapless states localized around the open edges of the system with nonzero bulk topological invariant. Recently, it was shown that the bulk-boundary correspondence does not work in non-Hermitian system \cite{nonh5,nonh6}. A question arises. Does it always work in Hermitian system? Although it is generally believed that it is true, there is no mathematical proof. A system with extended topological state, which breaks the bulk-boundary correspondence would be very interesting in the physics community.\\
In this Letter, we show that noninvertible Hamiltonians are topological. We discuss robustness conditions for the zero energy states. It is generally believed that topological edge states appear around edges where topological phase transition occurs. Here, we show that robust bulk zero energy states in a tight binding lattice can exist. We also explore some interesting systems that have robust states against disorder. 

\section{Formalism}
{\textit{ Consider a noninvertible Hamiltonian $\ds{\mathcal{H} }$ , i. e., the inverse Hamiltonian $\ds{\mathcal{H}^{-1} }$ does not exist.}} A Hamiltonian that can be represented by a $(N~{\times}~N)$ square matrix is noninvertible (singular) if and only if its determinant is equal to zero, $\ds{Det[\mathcal{H} ]=0}$.  {\textit{A noninvertible Hamiltonian has a zero eigenvalue with a nonzero eigenstate $\psi_0$, $\ds{\mathcal{H} \psi_0=0}$}}. In other words, the Hamiltonian $\ds{\mathcal{H} }$ is invertible if and only if 0 is not an eigenvalue of $\ds{\mathcal{H} }$. Let us now deform this noninvertible Hamiltonian adiabatically. {\textit{  The zero energy eigenvalue of the Hamiltonian $\ds{\mathcal{H} }$ is robust against such deformation as long as the  deformed Hamiltonian is noninvertible, too}}. This robustness is a signature of topological feature. We emphasize that nonzero energy eigenvalues of $\ds{\mathcal{H} }$ change with the deformation. We claim that the noninvertibility condition is the most general condition for the existence of robust zero-energy states and any zero-energy state in a topologically nontrivial system (with a nonzero Chern or $Z_2$ numbers, etc.) must automatically satisfy it. Our method have richer predictions than the standard theory of TI as will be shown below.  Note that our formalism is valid even for non-Hermitian systems. Below we will illustrate our idea in both Hermitian and non-Hermitian systems. The above approach is general and we need to define a new (topological) number for a systematic study of TI. We simply define this new number as the number of zero-energy states. Let $\ds{  \mathcal{H}^p  }$ be the powers of the $N{\times}N$ Hamiltonian, where $p=1,2,...$ are positive integers. It is easy to see that $\ds{  \mathcal{H}^p  }$ is noninvertible, too and the number of zero energy states of $\ds{  \mathcal{H}  }$ and $\ds{  \mathcal{H}^p  }$ are the same. However, their matrix ranks are not always the same since $\ds{    {   \textit{Rank}}(\mathcal{H}^p)   \leq  {   \textit{Rank}}(\mathcal{H}^{p^{\prime}})  }$ for $\ds{p>p^{\prime}}$ (They are the same for Hermitian Hamiltonians and not necessarily the same for non-Hermitian Hamiltonians). Consider the following set of the rank of these matrices, $\ds{{   \{  \textit{Rank}}(\mathcal{H}^p) \} }$. The minimum value of this set $\ds{{ \textit{Min} \{  \textit{Rank}}(\mathcal{H}^p) \}=r }$ determines the number of zero-energy states. It is given by $\ds{(N-r)}$ that is also the new number. It seems that this way of calculating zero energy states is not practical. Fortunately, we numerically check that it is enough to perform calculations for a few number of $p$ in most non-Hermitian systems. \\
A question arises. What is the physical origin of this phase? In a noninvertible matrix, at least one row (or one column) are linearly dependent on the others. From the physical point of view, this occurs if there is an order in the physical system. To be more specific, onsite potentials and tunneling amplitudes in a noninteracting system (interaction and exchange energies in an interacting system, etc.) should be arranged in such a way that some of them are linearly dependent on the others. A single-order in the whole system occurs if there is only one such linear dependence. The number of such orders is equal to the number of zero-energy states, which is also equal to the new number defined above. This is a new kind of order that is beyond the usual symmetry description. A phase transition occurs and the system becomes trivial if this order is broken. Below, we will give an example for a better understanding of this symmetry. \\
{\textit{A single-order in a tight binding lattice}: Consider a tight binding lattice with $N$ sites as shown in the Fig-1 for $N=8$. The lattice site number $n=1$ and $n=2$ have the same neighbors and the onsite potentials $V_n$ and tunneling amplitudes $t_{nm}$ for these two sites are symmetric. However, they are assumed to be completely disordered for the rest of the lattice sites and hence the system can be considered as an amorphous system \cite {amor}. Normally, we don't expect topological state in such a disordered system. Fortunately, there is a symmetry between the first two sites. As a result, a single-order occurs in the system and the first two rows of the corresponding Hamiltonian are the same. This implies that the Hamiltonian becomes noninvertible (its determinant is zero) and a zero-energy state appears in the system. We perform numerical computations and we see that the zero energy state is an edge state since the corresponding density is symmetrically localized on the sites $n=1,2$ and vanishes on the other sites. This zero energy edge state is robust against any kind of disorder or external potential provided that the single-order between the first two lattice sites is not broken. As an example, this zero energy state exists even in the presence of external electric (or gravitational) field as long as the first two sites are held at zero electric potential energy.\\
This example clearly shows that our approach simply predicts robust zero energy state without having to calculate Chern or any other topological number. Secondly, the symmetry considered here is different from the well-known discrete symmetries (particle-hole, chiral and time-reversal symmetries). The symmetry breaking here is also different from the usual Landau's symmetry breaking theory. Having discussed the topological single-order, let us now discuss another important issue. It is generally believed that band gap closing leads to topological phase transition. Furthermore, symmetry-protected robust modes appear around edges where topological phase transition occurs. We will show that these two statements are not always true as can be seen from the following example.
\begin{figure}[t]\label{2678ik0}
\includegraphics[width=8cm]{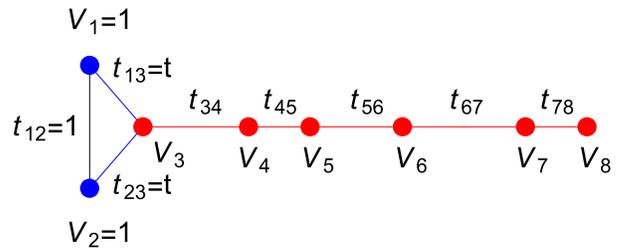}
\caption{ A tight-binding lattice with a topological single-order. The sites from $n=3$ to $n=8$ form a line while the first three sites make an isosceles (or equilateral) triangle. Here, $t_{nm}$ is the tunneling amplitudes between the sites $n$ and $m$, $V_n$ are the onsite potentials. $t_{nm}$ and $V_n$ take arbitrary values unless their values are specified in the figure. The topological single-order between the first two lattice sites occur as they have the same neighbor and the same onsite and symmetric tunneling amplitudes. Robust zero energy edge state appears in the system due to the topological single-order. The topological phase in the system is preserved as long as this single-order is not broken. }
\end{figure}\\
{\textit{Robust bulk states}}: Consider a  gapless tight binding lattice with an odd number of lattice sites, $2N+1$. The Hamiltonian reads $\ds{ \mathcal{H}(J)=\sum_{n=1}^{2N} J~(|n><n+1|+|n+1><n|)}$, where $J$ is the tunneling amplitude. Consider another tight binding lattice with an even number of lattice sites $2N$ and onsite potentials on the first two sites at the left edge: $\ds{ \mathcal{H}(J)=\sum_{n=1}^{2N-1} J~(|n><n+1|+|n+1><n|)  +}$ $\ds{J(|1><1|+|2><2|)   }$. These two systems have not staggered tunneling amplitudes as in the Su-Schrieffer-Heeger (SSH) model \cite{ssh} and are topologically trivial according to the standard theory of TI. However, both Hamiltonians are noninvertible, $\ds{Det[ \mathcal{H}(J) ]=0}$ and their topological numbers defined above are equal to $1$. Therefore there exists a zero-energy state in both systems. It is interesting to see that these zero-energy states are not localized around the edges. Instead they are extended states as shown in the Fig-2. This is in stark contrast to the bulk-boundary correspondence. Let us now study the robustness of these zero-energy states against tunneling amplitude disorder by introducing randomized tunneling amplitudes in the lattice: $J\rightarrow J_n= J+\delta_n$, where $\delta_n$ are site-dependent real-valued random set of constants in the interval $ [-\lambda,\lambda]$. We emphasize that disorder occurs at every lattice sites for the first system, while it is absent only at the left edge for the second one, i.e., $\delta_1=0$. We check that the disordered Hamiltonian is still noninvertible since $\ds{ Det[\mathcal{H}(J)]= Det[\mathcal{H} ( J+\delta_n)]  =0}$. This implies that zero-energy states are robust. We numerically see that they are robust even for very large values of $\lambda/J$.\\
This example shows that robust zero energy states have directly nothing to do with band-gap opening and closing. The only restriction is that the disorder should not break the noninvertibility of the total Hamiltonian for the robustness of zero energy states. Secondly, topological insulators have long been thought to be an insulator in the bulk and a conductor in its surface. Our findings challenge this view and find novel topological zero energy states in the bulk. \\
Suppose now that we change tunneling amplitudes in an alternating way to make these two systems gapped. According to the standard theory of TI, topological phase transition should occur because of the band gap opening. However, we see no such topological phase transition. Instead we see localized zero-energy states around the edges. Below, we illustrate our idea.
\begin{figure}[t]\label{2678ik0}
\includegraphics[width=3.85cm]{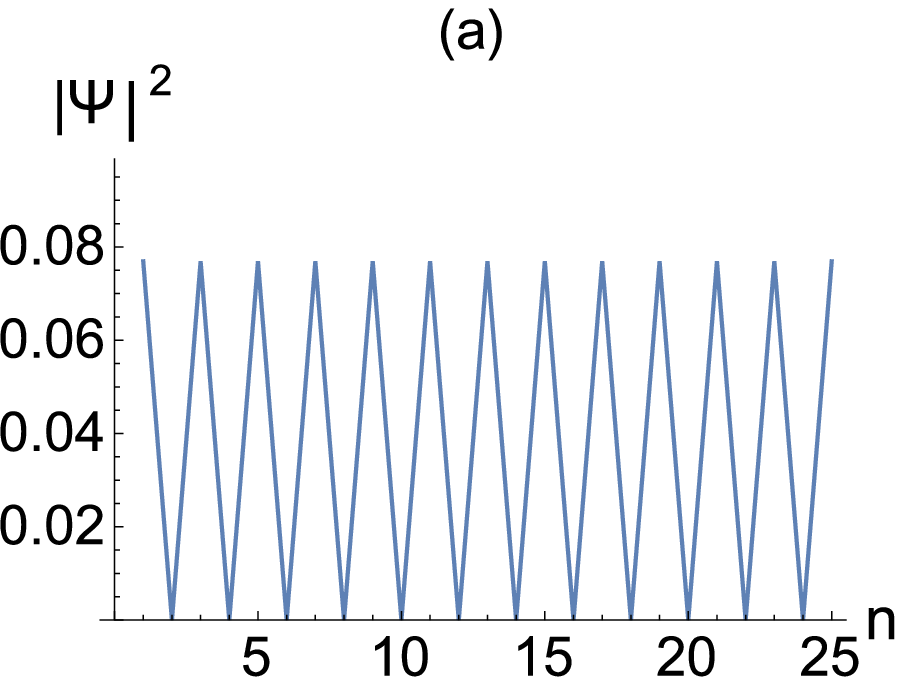}
\includegraphics[width=3.85cm]{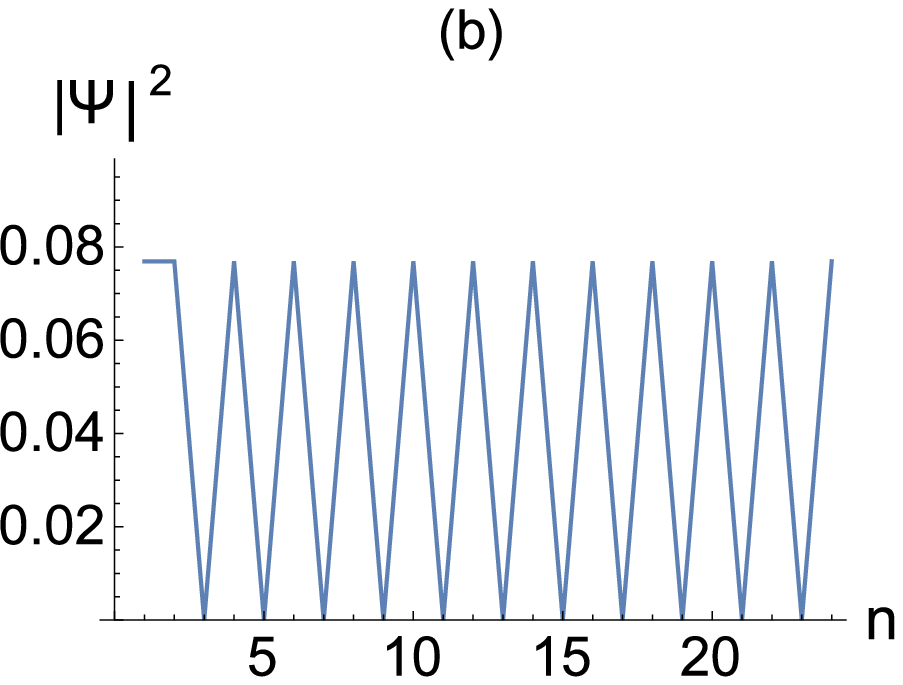}
\caption{ The densities of zero-energy states in the tight binding lattice with $J=1$. The number of lattice sites is odd for (a) with $2N+1=25$ and even for (b) with $2N=24$. The zero-energy states are not localized around the edges instead they are extended in the whole lattice. These states are robust against tunneling amplitude disorder as mentioned in the text.}
\end{figure}
\begin{figure}[t]\label{2678ik0}
\includegraphics[width=3.85cm]{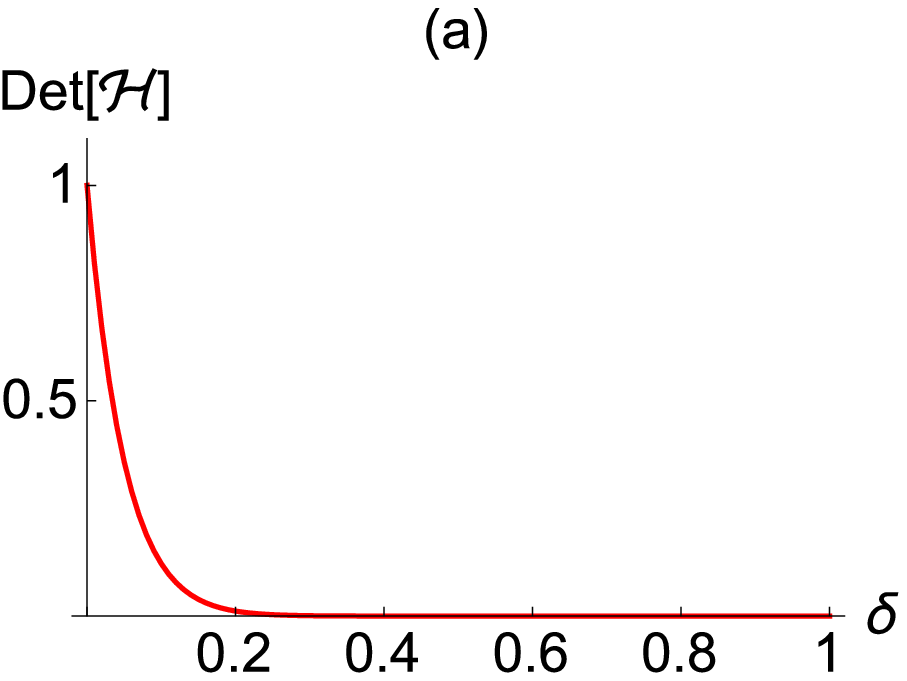}
\includegraphics[width=3.85cm]{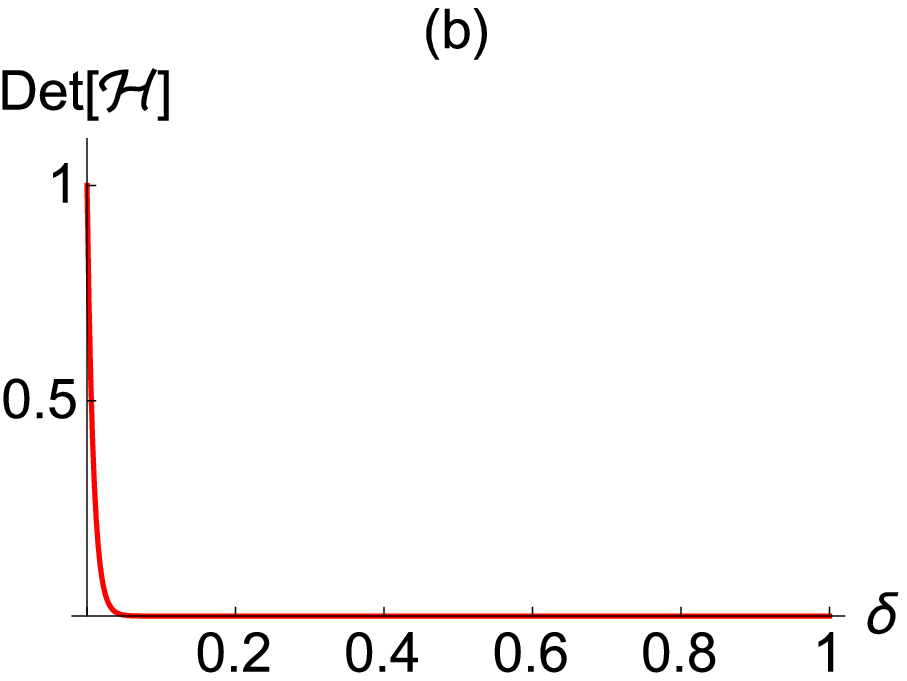}
\includegraphics[width=3.85cm]{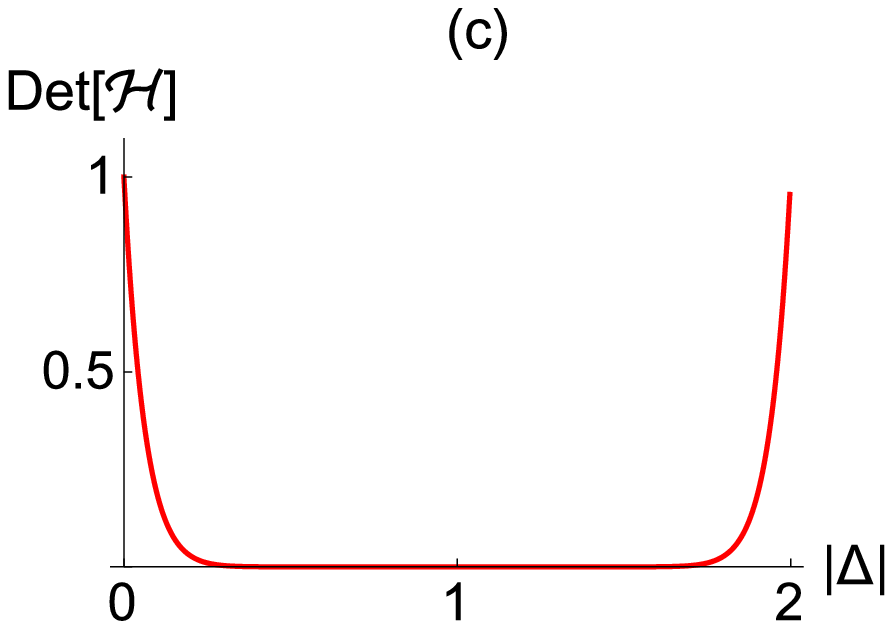}
\includegraphics[width=3.85cm]{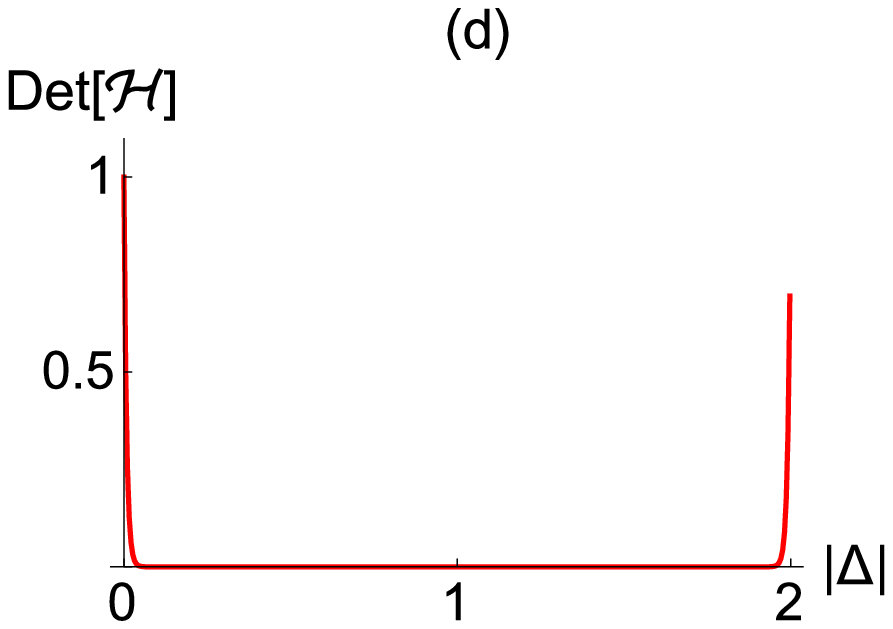}
\caption{ The determinants for the SSH model when $N=20$ (a) $N=120$ (b). In the limit $N\rightarrow \infty$, topological phase transition occurs at $\delta=0$. The lower panel is for the BHZ Hamiltonian at $A=1$ for $N=8$ (c) and $N=64$ (d). Note that the determinants are calculated at $k_x=0$ ($k_x\mp\pi$) when $\Delta<0$ ($\Delta>0$). As can be seen, $|\Delta|=2$ is the topological phase transition point for large values of $N$. }
\end{figure}\\
{\textit{The SSH model}: Consider now the SSH Hamiltonian \cite{ssh}: $\ds{ \mathcal{H}=\sum_{n=1}^{N-1} J_n~(|n><n+1|+|n+1><n|)}$, where the alternating tunneling amplitude is given by $J_n=1+(-1)^n\delta$ and $|\delta|<1$ is a constant. Let us first review the prediction of standard theory of TI. Imposing the periodical boundary condition on the system shows that topological phase transition occurs at $\delta=0$. The winding number is 1 for $\delta>0$ and 0 for $\delta>0$ and hence the system is topologically nontrivial (trivial) for $\delta>0$ ($\delta<0$). If the SSH lattice has two open edges, this prediction works if $N$ is large enough. In this case, zero-energy states appear around the edges. This is because of the fact that there is exponentially small overlap between the left and the right edge states. These overlapped states are not truly topological and not robust against the tunneling amplitude disorder. Let us now analyze this problem from our perspective. It is interesting to see that the determinant of the corresponding Hamiltonian is always zero for any value of $\delta$ when $N$ is odd. This implies that the system is always topologically nontrivial. The topological states are extended all over the latice at $\delta=0$ and they get localized around the edges if $\delta$ is increased. Conversely, $\delta=1$ is the topological phase transition point when $N$ is even (at this special value, the sites at both edges are separated from their neighbors) \cite{soz}. We find analytically that $\ds{  Det[\mathcal{H}]= (1-\delta)^{N}  }$, which increases very rapidly with $N$ when $~-1<\delta<0$. We see that $\delta=1$ is the exact topological phase transition point since the corresponding Hamiltonian becomes noninvertible at that point when $N$ is even. The system is not strictly topological for $\delta \neq1$. Fortunately, $\ds{  Det[\mathcal{H}]_{\delta>0} <<  Det[\mathcal{H}]_{\delta<0} }$ for large values of $N$. We numerically see that almost zero energy states appear when $0<\delta<1$ and $N$ is large. We plot the determinants of $\ds{\mathcal{H} }$ in the Fig-3.(a,b). The numerical computations show that our method explains the appearance of zero-energy states and their robustness very well and hence it is more powerful than the concept of the topological winding number, which does not distinguish whether $N$ is odd or even.  \\
We have applied our approach to the SSH model in $1D$. We stress that our approach works in any dimension. Let us now give another example in $2D$.\\
{\textit{Bernevig-Hughes-Zhang (BHZ) model }: Consider the BHZ model \cite{bernevig} $\ds{\mathcal{H} (\textbf{k} )=\left(\begin{array}{cc}   \mathcal{H}_1 (\textbf{k} )& 0 \\0 &  \mathcal{H}_1^{\star} (-\textbf{k} )\end{array}\right)}$, where $A$ and $\ds{\Delta}$ are real-valued parameters and  $   \mathcal{H}_1(\textbf{k} ) =A~(\sin k_x \sigma_x+\sin k_y \sigma_y)+\left(\Delta+\cos k_x +\cos k_y \right)\sigma_z  $. This system is a well-known $Z_2$ topological insulator with time-reversal and parity symmetries. \\
Consider a $2D$ strip that has $N$ sites along $y$-direction (open edge) and is infinite along $x$-direction. We calculate the corresponding determinant and find that it is zero when $-2<\Delta<0$ ($0<\Delta<2$) at $k_x=0$ ($k_x\mp \pi$). Therefore, the BHZ Hamiltonian is noninvertible and zero-energy states appear at these particular values of $k_x$. At different values of $k_x$, it is not noninvertible anymore and perfectly robust zero-energy states don't appear. However the determinant is not so big around these particular values of $k_x$ and hence almost robust nonzero energy states appear in the system (these are $`nearly ~robust$ ' states in the sense that their energy eigenvalues change slightly with the diosorder. Note that they can propagate without backscattering even if their eigenvalues changes slightly). For example, the determinants are equal to $3.1$ and $2.6\times 10^{15}$ at $k_x=\pi/12$ and $\pi/4$, respectively ($\ds{\Delta=-1.5}$ and $A=1$). Another interesting prediction is about the topological phase transition point. The BHZ system has topological $Z_2$ number $1$ for $|\Delta|<2$ and $0$ otherwise for $A=1$. (we assume that $\Delta\neq0$). We see that there is no such sharp transition point if the system has open edges. The number $N$ changes the topological phase transition point. If $N$ is large, then $|\Delta|=2$ becomes the transition point as predicted by the $Z_2$ number. However, this transition points change considerably for small values of $N$ (because of the wave function overlap between the two edges). This fact arises automatically in our formalism as can be seen from the Fig.3 (c,d). We define nearly topological states in the sense that their energy eigenvalues remain the same even in the presence of disorder. Although their energy eigenvalues change slightly with disorder, they can still propagate without backscattering. \\
So far, we have discussed that zero-energy states are perfectly robust. Invertible Hamiltonians whose determinants are close to zero can have almost robust states with nonzero energy eigenvalues. A question arises. Is there a perfectly robust state with nonzero-energy state?\\ 
{\textit{Robust non-zero energy states}: We are now in position to show that robust states with non-zero energy eigenvalues exist for invertible Hamiltonians.\\
Consider a noninvertible Hamiltonian $\mathcal{H}$. One can construct a new Hamiltonian $\ds{\mathcal{H} (m)=\mathcal{H}+m I}$, where $m$ is an arbitrary constant and $I$ is the identity matrix. This new Hamiltonian is invertible since $\ds{Det(\mathcal{H}+m I )\neq0}$. One can see that $\psi_0$ is the simultaneous eigenstate of both $\ds{\mathcal{H}}$ and $\ds{\mathcal{H} (m) }$ with eigenvalues zero and $m$, respectively: $\ds{\mathcal{H}\psi_0=0  }$ and $\ds{\mathcal{H}(m)\psi_0=m \psi_0 }$. Let us now add same disorder into the both systems. If the zero-energy state of $\mathcal{H}$ is robust against it, then nonzero-energy state of $\mathcal{H}(m)$ is robust, too. This is because of the fact that the deformed zero-energy state of $\ds{\mathcal{H} }$ is still the simultaneous eigenstate of $\ds{\mathcal{H}(m) }$ with eigenvalue $m$. We called this phase pseudo topological phase \cite{pseudo} as robust states still appear in a system which breaks chiral symmetry \cite{henrik1,henrik2,henrik3}.\\
To illustrate our idea consider the Hamiltonian $\ds{ \mathcal{H}=\sum_{n=1}^{N-1} J_n~(|n><n+1|+|n+1><n|)}$- $\ds{m \sum_{n=1}^{N}  (1+(-1)^n)(|n><n| )}$, where $J_n=1+(-1)^n\delta$. One can check that $\ds{Det[\mathcal{H} ]=0}$ if $N$ is odd. Therefore a topological zero-energy state appears in the system. This topological state is robust against the tunneling amplitude disorder as discussed above. Let us now construct $\ds{\mathcal{H} (m)=\mathcal{H}+m I}$ with $\ds{Det[\mathcal{H}(m) ]\neq0}$. This new Hamiltonian is in fact the massive SSH model \cite{hafi}. One can numerically see that the zero energy state of $\ds{\mathcal{H}}$ is simultaneous eigenstate of $\ds{\mathcal{H}(m)}$ with energy $m$ and this simultaneous state is robust in these two systems against tunneling amplitude disorder.  \\
{\textit{Non-Hermitian extension}: The first experimental demonstration of zero-energy states in a non-Hermitian system was made in 2017 \cite{nonh1} and non-Hermitian extension of topological phase is nowadays a hot topic. Our method doesn't distinguish Hermitian and non-Hermitian systems. Below, we study two interesting classes of non-Hermitian topological systems. \\
Suppose an antisymmetric Hamiltonian, $\ds{\mathcal{H} =-{\mathcal{H} ^T}}$, where $\ds{\mathcal{H}^T}$ is the matrix transpose. Note that an antisymmetric Hamiltonian is non-Hermitian. All antisymmetric matrices of odd dimension are automatically noninvertible. Therefore we get a huge family of non-Hermitian Hamiltonians with zero-energy states. They are automatically robust against disorder that respects the antisymmetric character of the non-Hermitian Hamiltonian. We stress that they may also be robust against some other kind of disorder depending on the specific Hamiltonians.\\
Let us now study another class. Suppose that $\ds{\mathcal{H} }$ has a row (or column) that is all zeros. Then $\ds{\mathcal{H} }$ is a noninvertible matrix and its determinant is zero. This is possible only for non-Hermitian Hamiltonians with asymmetric tunneling. Assume further that the matrix form of disordered Hamiltonian has also the same $0$ row (or $0$ column). Then the disordered Hamiltonian is noninvertible and zero energy states are robust.\\
Non-Hermitian topological systems are not restricted to the above two systems and one can find many interesting and experimentally realizable non-Hermitian topological systems using our formalism.\\
{\textit{Symmetry protection}:
Topological modes are known to be protected by some symmetries. We will explore discrete symmetry properties of our system. \\
Consider firstly a chiral symmetric Hamiltonian: $\ds{ \mathcal{S}    \mathcal{H}   \mathcal{S}^{-1} = - \mathcal{H}     }$, where $\ds{ \mathcal{S}  }$ is an {\textit{invertible chiral}} operator. The determinant of such a Hamiltonian when $N$ is an odd number is always zero. Let us now deform the Hamiltonian in such a way that the chiral symmetry remains intact (The deforming potential has the chiral symmetry, too). Note that the addition of two noninvertible matrices is not necessarily noninvertible. Fortunately, the deformed Hamiltonian has the same chiral symmetry and hence it is still noninvertible. As a result, one can say that robust zero-energy states protected by the chiral symmetry exist for $\ds{ \mathcal{H}  }$ if $N$ is an odd number. We can generalize this formalism to time-reversal $\ds{ \mathcal{T}  }$ and particle-hole $\ds{ \mathcal{C}  }$ symmetric systems with open edges.  \\
In conclusion, we have shown that noninvertible Hamiltonians are topological  and have zero energy states. We have also shown that noninvertibility condition determines exact topological phase transition point in the 2D BHZ model. We have discussed that zero-energy states are robust against any kind of deformation as long as the deformed Hamiltonian is noninvertible, too. We have predicted robust zero-energy bulk states in a tight binding lattice. We emphasize that our method can be applied to interacting systems, too. For example, one can study topological phase for the Hubbard model. As a final remark, topological superconductors with Majorana zero modes and topological semimetals can also be studied using our formalism. We think our study will pave the way for the understanding of topological systems.


\begin{thebibliography}{0}
\bibitem{soz} To see this clearly, consider the SSH Hamiltonian for $N=3$ and $N=4$. If $N=3$, the first and the third rows are linearly dependent to each other. Therefore, the rank of the corresponding matrix is $2$. However, no such linear dependence can be found for $N=4$ unless $\delta=1$.
\bibitem{hafi} If we impose periodical boundary conditions. it becomes the massive SSH model: $\ds{\mathcal{H}_{\beta}=\left(J_1+J_2\cos(k) \right)\sigma_x+J_2\sin(k)\sigma_y+m\sigma_z}$, where $\vec{\sigma}$ are Pauli matrices, the crystal momentum $k$ runs over the first Brillouin zone and $-\pi<k<\pi$.
\end{thebibliography}
\end{document}